# Defense via Behavior Attestation against Attacks in Connected and Automated Vehicles based Federated Learning Systems

Godwin Badu-Marfo[1], Ranwa Al Mallah[2], Bilal Farooq[1]

*Abstract*— The recent application of Federated Learning algorithms in IOT and Wireless vehicular networks have given rise to newer cyber threats in the mobile environment which hitherto were not present in traditional fixed networks. These threats arise due to the intrinsic nature of wireless transmission medium and other inherent characteristics of mobile networks such as high-node mobility and rapidly changing topology. This paper investigates the robustness of Vehicular AttestedFL defense strategies against falsified information attacks by tracking the behavior. We show that the defense strategies are capable of detecting and eliminating malicious nodes in the wireless mobile setting of the future smart road networks.

## I. INTRODUCTION

Learning models continuously improve their accuracy on prediction by updating gradients through collaborative training on a wealth of data gathered from pervasive data sources. Generally, open data sharing is remarkable but could suffer data privacy leakages and security threats for the data owners [1]. To address this privacy concern, McMahan et al. [2] proposed "Federated Learning" (FL), a decentralized training architecture for shared model training without data leaving the device of the end user.

Federated Learning [2] allows for learning smarter models over lower network latency while using less power consumption on heterogeneous devices, all while ensuring privacy. The FL process typically involves a chief node who acts as an aggregator of model weights received from a set of computing nodes (workers) that independently train a global model on their local data without the data leaving the devices. The Chief periodically receives model weights from the workers after a training round then performs a weighted average on the received model parameters to derive a global model which is sent back to the workers.

Federated Learning is a growing research domain and its application has been considered in data-driven fields which demand personally identifiable information especially health, finance and many others. In the context of connected and vehicular networks, FL exhibits the capacity of providing a privacy-by-design framework to harness the full potential of intelligent transportation systems coordinated by a network of infrastructure units that collaboratively exchange information to create safer roads. Liu et al.[3] propose a FedGRU algorithm for traffic flow prediction with federated learning for privacy protection by using a secure parameter aggregation mechanism to train a global model in a distributed manner.

Cooperation and coordination among CAVs is enabled through Vehicle-to-Vehicle (V2V) and Vehicle-to-Infrastructure (V2I) wireless communications. Vehicles exchange messages with each other and communicate with their associated Roadside Unit (RSU). However, the broadcast of vehicular information through wireless networks could impose threats on data leakages and security breaches. Similarly, adversaries could gain illegal access to connecting devices to embark on a denial-of-service (DOS) attack or poisoning attack [4] which sends falsified or fabricated messages across the network. In poisoning attacks, both the learning model and training data are susceptible to being falsified. In this regard, when adversarial attacks impact on model weights such that local updates sent to the chief are manipulated, this is referred as "*model poisoning attack*".

Typically, model poisoning attacks occur in the training phase of the model. The adversary aims to perturb the weight parameters of the local model such that the performance and prediction accuracy of the global model are degraded [5]. Poisoning attacks are broadly classified into two; *untargeted* or *targeted* attacks. The objective of the untargeted poisoning attack (i.e., Byzantine attack) is to reduce the accuracy of the model prediction for all output classes. The attacker's aim is to diverge the model parameters to the point where they cannot be trained without resetting to a previous checkpoint hence reducing the test accuracy of the model [6]. On the contrary, the targeted attack aims to modify the model weights such that specific output classes chosen by the attacker are misclassified [7]. Generally, vulnerabilities evolve in a dynamic and fast growing environment (i.e. FL) hence this arises the need to design new defense strategies to meet adversarial attacks. For example, the work of [8] et al. shown that a direct application of the FL process without any consideration of the underlying communication infrastructure of the CAVs could expose the online training of models to cyberattacks. The authors proved that a malicious entity could exploit vulnerabilities in the vehicular network in order to poison the training of the model with false inputs. In this regard, this paper aims to expand on research findings in the protection against adversarial poisoning attacks by exploiting new defense mechanism that guarantee protection against poisoning attacks. We experiment with a novel detection and behavioural pattern analysis of defense mechanism, the *attestedFL* [9] that promises of robustness in adversarial settings through monitoring of the behavior of the worker nodes over time on their state persistence and removes unreliable nodes from the aggregation process. In order to test the robustness of this defence mechanism, we conduct untargeted attacks on various federated learning settings and present different attacking patterns. Subsequently, we evaluate the impact of the attestedFL defense by observing the prediction accuracy over each epoch.

[1]Godwin Badu-Marfo and Bilal Farooq are with the Laboratory of Innovations in Transportation, Toronto Metropolitan University, ON, Canada gbmarfo@torontomu.ca; bilal.farooq@torontomu.ca

[2]Ranwa Al Mallah is with the Computer Security Lab, Royal Military College of Canada, ON, Canada ranwa.al-mallah@rmc-cmr.ca

The contributions of this paper are summarized as follows:

- We propose a framework adapted for FL based systems in vehicular networks and on which a defense mechanism is able to run to ensure protection against poisoning attacks. The framework enables the exchange of the underling temporal and dynamic local model updates of every vehicle of the system in a transparent and secure way for the purpose of monitoring their training.
- We propose Vehicular AttestedFL consisting of three lines of defenses suitable in vehicular networks where the mobility of the vehicles is taken into account.
- We implement attacks on a realistic federated learning task in vehicular networks by developing a predictive model for link level speed to predict speed of road segments and evaluate the impact of the defense in order to validate the efficiency and security of the protection mechanism.

The rest of the paper is organised as follows. In Section II we present a brief literature review. In Section III we describe the threat model and illustrate the attacks. We present the framework and Vehicular AttestedFL in Section IV. In Section V, we provide experimental results and analysis of the impact of the defense against poisoning attacks in vehicular networks. We conclude the paper and provide future work in Section VI.

## II. LITERATURE REVIEW

### A. Federated Learning in Vehicular Network

Federated Learning [10] has evolved as a privacy-by-design learning technique having the capability of collaborative learning without training data points leaving the device of the data owner. Unlike traditional machine learning methods where a single node learns the best fitting parameters to explain variances in training data points, federated learning works in a distributed manner by delegating the learning process to multiple client nodes who independently train a global model on their data points to achieve local parameter updates, then subsequently shares the respective local model updates to a chief node that acts as an aggregator to compute the mean weighted global model. The global model updates is iterated on a number of epochs until the global model reaches convergence in its prediction accuracy. Federated learning has gained dominance in academic literature and has been adopted for numerous application areas. For example, the work of Nikman et al. [11] discussed numerous applications of FL in the wireless networks, especially in the context of 5G networks. As an approach to reduce backhaul traffic load, content caching and data computing is implemented at the edge nodes of the wireless network. FL leverages on gradients of locally trained models to train a centralized server (i.e. Chief) without direct access to the user data as applied in content popularity prediction in proactive caching in wireless networks. Similarly, FL has been used in the wireless networks to learn the activities of mobile phone users. This application implements a search mechanism for information retrieval or in-app navigation. Also, Bonawitz et al. [12] proposed a production-level FL architecture to focus primarily on the averaging algorithm running on mobile phones that conceived an environment of mobile devices with lower bandwidth and reliability when compared to data center nodes. The work of Nishio et al. [13] applied FL in different environmental conditions, example in scenarios where the Chief node can reach any subset of devices to initiate a training round, but receives model updates sequentially due to cellular bandwidth limit.

In the context of vehicular networks, FL is adapted in predicting the real-time state of traffic conditions with CAVs. As an example, a fleet of Autonomous Vehicles may require an updated prediction model of traffic, construction zone delays, or pedestrian behavior to operate safely on the roads. FL can help to train models that efficiently adapt to changes in these situations, while maintaining user privacy. Samarakoon et al. [14] proposed a FL model in the context of V2V communication to learn the distribution of the extreme events related to queue delays. Similarly, Lu et al. [15] presented a work on demonstrating the effectiveness of driver personalization in connected vehicles to predict failures in an effort to ensure sustainable and reliable dirving in a collaborative fashion using federated learning and long short term memory (LSTM) networks. Shiva et al. [16] proposed a communication efficient and privacy preserving federated learning framework for enhancing the TCP Performance over WiFi of Internet of Vehicles.

In a vehicular network, we execute FL protocol by considering zonal unit as the *chief* with a number of RSUs acting as transient units through which model parameters are sent from the vehicles to the chief with vehicles acting as *workers* as shown in Fig.1. The FL process is executed as follows:

- The chief considers the available workers or selects subset and broadcast the model weights to them as workers are expected to stay connected to the chief for the duration of the round.
- Upon receiving the weights, each worker is expected to perform local model computation and broadcast the model back to chief.
- As local model updates are received, the chief computes the weights for the next round by aggregating the updates using federated averaging algorithm (FedAvg).
- The process is repeated with the addition of more workers and the round goes on until a desired performance or model accuracy is achieved by the chief.

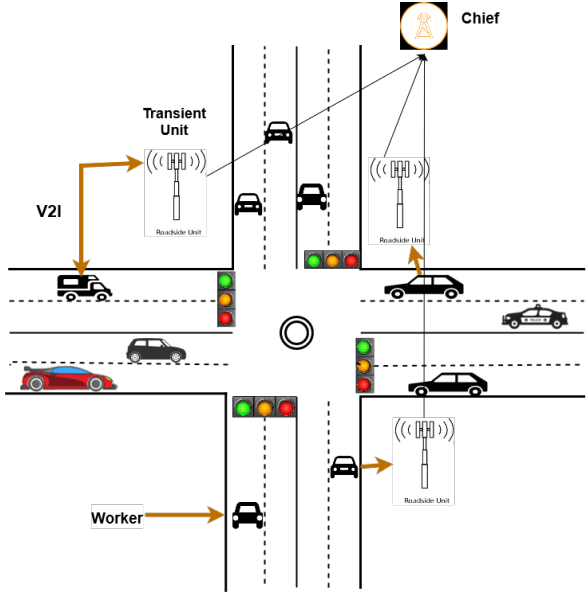

Fig. 1. Vehicles acting as workers in the FL process communicate using V2I with the chief implemented at the Road zonal master

*B. Defenses against Poisoning Attacks in Vehicular Networks*

The goal of attack detection methods is to distinguish malicious activities from the regular patterns in intelligent networks. To mitigate various variant poisoning attacks, detection methods must be achieved to ensure that there is no security threats of the data collected by the devices. There are poisoning attacks that involves identification of unusual features during training in the FL process.The adversarial goal of the poisoning attacks is to affect the convergence of the global model because of the malicious local model updates that the attacker may send back to *chief* node in the FL process [17]. Hence, there is the need to put in defense mechanisms against such attacks. Several detection and defense mechanisms against poisonous attacks in FL proposed by [6] [18] [7] aim at ensuring convergence of the global model. Recently, federated learning-based defenses against poisonous attacks in vehicular networks have been proposed. Liu et al. [19] propose a blockchain and federated learning for collaborative intrusion detection for vehicular edge computing networks and achieved the final higher accuracy model aggregation and sharing through multi-party aggregation training. In their defense strategy, two stage intrusion detection systems (IDS) and trust-based incentive mechanism are designed by leveraging FL through multiple edge vehicles and RSUs that collaborate together. The distributed model aggregation scheme based on the blockchain ensures the security of the model sharing and storage. Moreover, Qi et al. [20] design a blockchain-based secure FL framework for urban traffic flow management by introducing an FL framework to protect the privacy of the vehicle data and defend against poisoning attacks. Specifically, they leverage blockchain to implement a decentralized FL framework to achieve accurate model prediction whilst protecting privacy in vehicle location sharing. These defense strategies help defend against poisoning attacks in FL by incorporating blockchain technology. However, to the best of our knowledge, no previous work has studied the behavioural-based defenses against poisoning attack on FL based systems in CAVs.

## III. THREAT MODEL

The threat model of model poisoning attacks is related to the attacker's knowledge, the attacker's goal, the attacker's strategy, and the attacker's ability to influence the training data. Vehicles are able to continuously extract link level information such as images of their surroundings and use them locally to participate, in a distributed manner, in the training of a global model. Vehicles travel the zone under the coverage of the RSU and then broadcast their local model update to the RSU. Due to privacy constraints, the vehicles in the network do not need to transmit precise location data such as GPS positions to the RSU. The system only needs to be informed of the vehicle IDs under the coverage of an RSU but not the precise location of each vehicle. We assume that the attacker has partial knowledge of the undergoing FL process. The attacker does not need to have information about the model parameters or algorithms running at the chief node. Also, before attacking the distributed training, the attacker does not need to perform reconnaissance by studying the coverage area of the RSU, the appropriate timing to perform the attacks, the number of malicious inputs to inject and the duration of the attack. We only assume that the attacker has the ability to compromise the on-board unit (OBU) of the vehicle in order to transmit malicious messages. This can be performed physically, wirelessly, or via malware [21]. The attacker's goal is to sabotage the ongoing FL process and indirectly influence the parameters of the learned model by adding malicious updates to the chief node as in [8]. In this paper, we focus on how the Vehicular AttestedFL is designed to defend against the falsified information and Sybil poisoning attacks in FL classification task. In those poisoning attacks, compromised information is sent by the malicious vehicle that is moving in and out of the zone under study very rapidly and thus continuously providing falsified real-time updates to the RSU. The zone under study is the area where the RSUs act as transient units that receive messages from the connected vehicles and transmit to the zonal master.

## IV. VEHICULAR ATTESTEDFL DESIGN

We propose a framework adapted for FL based systems in vehicular networks and on which a defense mechanism is able to run to ensure protection against poisoning attacks. In Ranwa et al. [8] the authors considered only one RSU serving as the chief node that receives model parameters from the OBU of all the vehicles within the zone under study. However, in our study, we develop a network FL which does not consider an RSU as the chief but considers a number of RSUs acting as transitory unit that receive model weights and sends them to the zonal unit for global model aggregation. In this scenario, although the RSUs have a computing capacity they do not act as chief node to do model aggregation but act as transient unit. Thus, the zonal unit acts as the chief node that performs the model aggregation.

This defence strategy protects against falsified information and Sybil attacks whose main objective is to prevent convergence of the global model. Whilst falsified information attacks seek to substantially degrade the performance of the FL process that rely on averaging to generate the global model Sybil attackers are only interested in producing a larger attack impact and at the same time avoid detection [8]. The framework aims at assessing if the worker is reliable by observing if the node is really training iteration after iteration. Unreliable workers act as attackers that are not training and are neutralized. The detailed design of Vehicular AttestedFL consists of three components of defense that protects against local model poisoning attacks.

---

**Algorithm 1** Vehicular AttestedFL Algorithm

**Input:** Set of road segments under the coverage of the RSU, RSc, Global Model, $GM^t$, parameters sent by the chief sent at iteration t to the workers in its coverage area, Local Model updates $LM_{i,t+1}$ of each *worker i*, $H_{i,z}$ a subset $z$ of a *worker's* previously uploaded consecutive Local Model update recorded as a pair of $LM$ and $GM$ at that time

**Output:** Reliable Global Model $GM_{t+1}$ at iteration $t$

**Function** Main():

**for** *for iteration t* **do**
    Reliable = false;
    **for** *for n workers i* **do**
        Let St be the weight of indicative features at iteration $t$;
        **for** *n Hi,z* **do**
            **if** *Conditions of vehicular AttestedFL-1* **then**
                **if** *Conditions of vehicular AttestedFL-2* **then**
                      **if** *Conditions of Vehicular AttestedFL-3* **then**
                          Reliable = true ;
                      **end**
                **end**
            **end**
        **if** *Reliable = True* **then** ;;
        Let *Real* be the vector containing the index i of all reliable workers;
    **end**
    Federated aggregation of the $LM_{i,t+1}$ Real *workers* in Real;
**end**

---

- **Vehicular AttestedFL1:** This line of defense seeks to analyze the history of the workers' updated local models and the convergence of the local model updates towards the global model in order to eliminate non-training local models. Whilst the Euclidean distance increases during model training in targeted attacks, there is no increase or decrease in the Euclidean distance during the whole training phase in comparison to the global model. There is an assumption that the mean and the standard deviation of the Euclidean distance of the model of a benign worker at a given iteration becomes less than the Euclidean distance of a malicious worker making the convergence speed of the benign nodes faster as compared to the malicious nodes.

- **Vehicular AttestedFL2:** This involves measurement of the cosine similarity of successive local model updates and the behaviour of the angular distance throughout the training of a node to remove abnormal training behavior and discard unreliable nodes from the aggregation process. A worker's node is considered unreliable if it does not show correlation over time between its local model updates. For nodes of benign workers, as training progresses, the cosine similarity metric indicates the similarity of the successive local model, LMj updates. On the other hand, the score of cosine similarity between local model updates of a single worker continuously decreases as training progresses in the case of untargeted poisoning attacks.

- **Vehicular AttestedFL3:** Similarly, Vehicular AttestedFL 3 measures reliability of workers based on the assumption that the chief has a small validation dataset as the chief can test to see how the local model of a worker predicts on the validation dataset. This involves measuring the performance of the same sample over a set of iterations within a time frame in order to monitor performance of each work and compare it to previous performances on that validation dataset. In this scenario, if a node trains better on a validation set the performance improves and it is considered reliable whereas a node is considered unreliable if its error rate increases over time due to the fact that attackers are unconcerned in the local model training.

## V. EVALUATION

We used the road network in Toronto to set up the evaluation.

### A. Case study and training dataset

For the implementation of the defense strategies, we consider a real-world scenario for our experiment. The speed data collected from road links in Toronto Downtown using speed recorders at refresh rate of five (5) minutes were used for this experiment. The road network covers seventy-six (76) intersections and two-hundred and twenty-three route links. While the training data represents the traffic speed of the study region over time, we use this historical data, $v_{t-M+1}$ to $v_t$ to predict the speed at a future timestep, $v_{t+H}$, where $H$ means the next $H$ time steps and $M$ means the previous $M$ traffic observations ($v_{t-M+1}$, , $v_t$).

In this experiment, we simulate our proposed FL defense mechanism on five (5) Roadside Units (RSU) that act as worker nodes and one (1) chief node. Training dataset is sharded into smaller dataset and distributed onto the worker nodes.

## B. Experimental Setup

To demonstrate the untargeted model poisoning attack, an FL process is implemented which involves training Long Short Term Memory (LSTM) deep neural network to predict speed on a link as in [8]. The model implements a simplified network composed of a single LSTM layer and a fully connected network having a regressing Linear output layer. Every worker node implements the same model network architecture but trains on its independent datasets using stochastic gradient descent as optimizer.

The FL defense algorithm is implemented in python using deep learning framework, *PyTorch*. There is random selection of workers in each round and the computed SGD updates are sent to the chief node for aggregation. The attackers are selected from five (5) workers in each communication round. We simulate the two types of attacks, Sybil attacks and falsified information attacks and evaluate their performance of our defense when the FL process is under attack. The model accuracy is evaluated on the testing data.

We conduct three experiments to simulate the robustness of our defense mechanisms against three (3) poisoning attack models; static, pretence and randomized attacks. In the static attack, one worker node is elected as an attacker who consistently sends random weights as its model output. For the pretence attack, two (2) worker nodes are elected as attackers and randomly sends random weights or trained weights.

## C. Results

In this section, the results of the defense-attack simulations are discussed. We discuss the analysis of Federated Learning with Poisoining attack (one of the five nodes sends random noise rather than observed speed data) in three scenarios of static attack (continously emit random noise), pretence attack (attacker acts benign or malicious at specific points in training) and randomized attack (either nodes could randomly be malicious).

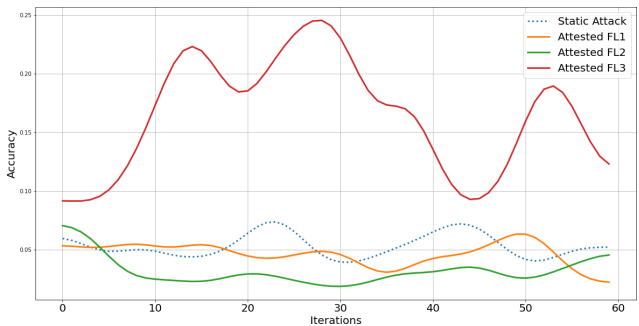

Fig. 2. Attested FL defenses on static poisoning attacks

The performance of Vehicular Attested FL1 presented in Figure 2 gives a satisfactory and an effective defense mechanism that is capable of eliminating non-training local models by studying the deviation from the global model with euclidean distances of the worker model updates to the global model. We assume the malicious worker node does not learn hence its parameters are not updated during the training phase. In effect, the euclidean distances are widened from the global model using a threshold value. Vehicular Attested FL1 proves its robustness by improving the detection accuracy of such malicious nodes and eliminating them from federated averaging at the chief and that impacts the convergence of the global model evident by the consistent decays of validation loss on multiple iterations as shown in Figure 2. Similarly, Vehicular Attested FL2 also gives satisfactory results in defending against static attacks. Using the cosine similarity between successive model updates as the algorithm for detecting malicious nodes, the Attested FL2 algorithm presents a stronger and better robust defense mechanism for static attacks. This is evident by the smooth decay of validation loss observed at the early stage of training given credence to its ability to detect malicious nodes at early onset of federated training. For the Attested FL1, the validation losses are generally stagnant for the early stage of training but get better in detection when multiple iterations are done. However, the Attested FL2 presents a faster detection and elimination of malicious nodes which does not require long span training iterations to improve its performance. The results of Vehicular AttestedFL3 present an unsatisfactory and ineffective defense mechanism for static poisoning attacks. The algorithm fails to provide a consistent decay of validation loss on the global model and generally presents a non-linear performance outlook on static poisoning attacks suggesting a weak strategy. While we assume model weights of the poisoning attack from a random normal distribution, the mean and standard deviations are generally equivalent and consistent for successive sampling hence the error rate of prediction could vary in tandem to deviations of the model weights sampling. In effect, attestedFL 3 is unable to flag and detect insignificant variations in error rate of prediction on successive model weights.

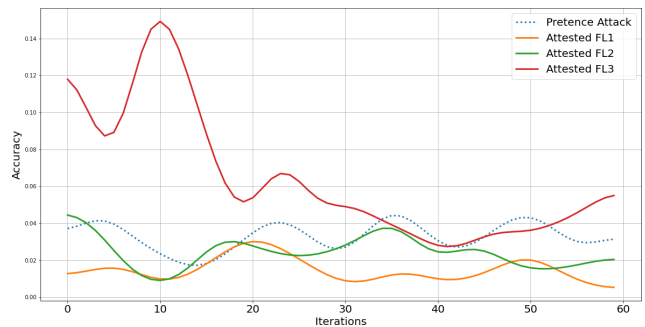

Fig. 3. Attested FL defenses on static pretence attacks

In scenario 2 of attested FL defenses on static pretense attack, the single node acts as an attacker who trains and updates its model weights for a specified rounds of iteration then subsequently switches to a non-training and malicious mode where it sends falsified model weights to the chief. The

| Attacks | AFL1 | AFL2 | AFL3 |
|---|---|---|---|
| Static Attack | Good | Strong | Weak |
| Static Pretence Attack | Strong | Weak | Good |
| Randomized Pretence Attack | Strong | Weak | Good |

results of robustness performance of AttestedFL defenses are shown in Figure 3. AttestedFL1 using the euclidean distance of successive weights as the baseline for detection, presents the best satisfactory performance showing lower validation losses that steadily decay over multiple iterations to achieve convergence. Similarly, AttestedFL2 also performs weakly with lower losses below the benchmarked validation losses of the attacker. However, a lot of incremental losses are witnessed suggesting model training instability hence making convergence of the global model intractable. However, AttestedFL3 showed impressive successes with detecting and eliminating pretentious nodes that periodically ingested falsified weights. This is represented by the steady drop in validation losses on the global model suggesting optimal learning of the global model. Though the losses of AttestedFL3 are quite higher, this could be improved with longer training epochs.

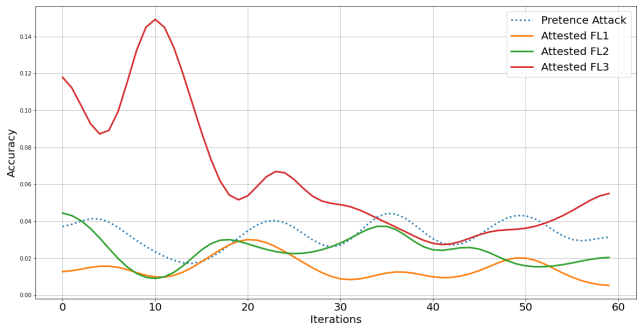

Fig. 4. Attested FL defenses on randomized pretence attacks

In scenario 3 of attested FL defenses on randomized pretense attack, nodes are randomly selected to act as a pretentious attacker that trains and updates model weights for a specified rounds of iteration then subsequently switches to a non-training and malicious mode where it sends falsified model weights to the chief. Similar to the results of the static pretense attack, the results of the randomized pretense attacks shown in Figure 4 has Vehicular AttestedFL1 presenting the best robustness performance with lower validation losses that decay over multiple iterations to achieve convergence. Also, Vehicular AttestedFL2 performs weakly with lower losses below the benchmarked validation losses of the attacker. An oscillation of incremental losses is observed and that suggests model training instability which could make achieving convergence difficult. Vehicular AttestedFL3 gave rather distinctive robustness performance with detecting and eliminating pretentious nodes that periodically ingested falsified weights. This is shown by the gradual drop in validation losses on the global model suggesting optimal learning of the global model. Though the losses of Vehicular AttestedFL3 are quite higher, this could be improved with longer training epochs.

## VI. CONCLUSION

In this work, we have evaluated the robustness of the attestedFL defenses against sybil attack vulnerability of FL in the vehicular networks. These defense mechanisms are urgently required since mobility networks are largely susceptible to adversarial attacks. Through this experimentation, the Vehicular AttestedFL1 algorithm of euclidean distance provided an optimal defense guarantee across the attack simulations. While a specified threshold is set to elect malicious workers on their successive euclidean distances, the best threshold value still remains an open question. Future work will investigate the optimal threshold values and also perform a thorough sensitivity analysis on the parameters to achieve robustness against CAV sybil attackers.